\documentclass[journal=nalefd,manuscript=letter,layout=twocolumn]{achemso}

\usepackage{amsmath,amssymb,amsfonts}
\usepackage{newtxtext,newtxmath}
\usepackage[utf8]{inputenx}
\usepackage{graphicx}
\usepackage[dvipsnames]{xcolor}
\usepackage{cuted}
\usepackage{xspace}
\usepackage{soul} 

\usepackage[pdftex,unicode=true,bookmarks=false,breaklinks=false,pdfborder={0 0 1},colorlinks=true]{hyperref}
\hypersetup{linkcolor=blue,citecolor=blue,urlcolor=blue}

\newcommand{\celsius}{$~^{\circ}$C\xspace}
\newcommand{\ingaas}{In$_x$Ga$_{1-x}$As\xspace}
\newcommand{\inShell}{In$_{0.15}$Ga$_{0.85}$As\xspace}

\author{Hanno Küpers}
\author{Ryan B. Lewis}
\affiliation{Paul-Drude-Institut für Festkörperelektronik, Leibniz-Institut im Forschungsverbund Berlin e.V., Hausvogteiplatz 5--7, 10117 Berlin, Germany}
\alsoaffiliation{Present address: Department of Engineering Physics, McMaster University, L8S 4L7 Hamilton, Canada}
\author{Pierre Corfdir}
\affiliation{Paul-Drude-Institut für Festkörperelektronik, Leibniz-Institut im Forschungsverbund Berlin e.V., Hausvogteiplatz 5--7, 10117 Berlin, Germany}
\alsoaffiliation{Present address: ABB Corporate Research, 5405 Baden-Dättwil, Switzerland}
\author{Michael Niehle}
\author{Timur Flissikowski}
\author{Holger T. Grahn}
\author{Achim Trampert}
\author{Oliver Brandt}
\author{Lutz Geelhaar}
\affiliation{Paul-Drude-Institut für Festkörperelektronik, Leibniz-Institut im Forschungsverbund Berlin e.V., Hausvogteiplatz 5--7, 10117 Berlin, Germany}
\email{geelhaar@pdi-berlin.de}

\title{Drastic effect of sequential deposition resulting from flux directionality on the luminescence efficiency of nanowire shells}


\begin{document}
\graphicspath{{./figures/}}

\begin{abstract}
Core-shell nanowire heterostructures form the basis for many innovative
devices. When compound nanowire shells are grown by directional
deposition techniques, the azimuthal position of the sources for the
different constituents in the growth reactor, substrate rotation, and
nanowire self-shadowing inevitably lead to sequential deposition. Here,
we uncover for \inShell/GaAs shell quantum wells grown by molecular beam
epitaxy a drastic impact of this sequentiality on the luminescence
efficiency. The photoluminescence intensity of shell quantum wells grown
with a flux sequence corresponding to migration enhanced epitaxy, \textit{i.\,e.}
when As and the group-III metals essentially do not impinge at the same
time, is more than two orders of magnitude higher than for shell quantum
wells prepared with substantially overlapping fluxes. Transmission
electron microscopy does not reveal any extended defects explaining this
difference. Our analysis of photoluminescence transients shows that
co-deposition has two detrimental microscopic effects. First, a higher
density of electrically active point defects leads to internal electric
fields reducing the electron-hole wave function overlap. Second, more point defects
form that act as nonradiative recombination centers. Our study
demonstrates that the source arrangement of the growth reactor, which is
of mere technical relevance for planar structures, can have drastic
consequences for the materials properties of nanowire shells. We expect
that this finding holds also for other alloy nanowire shells.
\end{abstract}


\section*{Main text}

Semiconductor nanowires (NWs) offer exciting opportunities for the fabrication of electronic and optoelectronic devices with improved efficiencies and functionalities compared to their planar counterparts\cite{garnett2019}. In many cases, the functionality arises from a shell grown around a NW core\cite{royo2017}. Shell growth can be employed for the formation of heterostructures that underlie essentially all semiconductor devices and can lead, most eminently, to quantum confinement. By growing heterostructures in a NW core-shell geometry, the high surface-to-volume ratio of such structures can be exploited to enlarge the effective area compared to planar layers\cite{waag2011}. Furthermore, in such structures the strain is partitioned between the core and shell, thus enabling the combination of highly lattice-mismatched materials\cite{Kavanagh2010}. Concrete examples of devices and applications based on core-shell NWs include solar cells\cite{czaban2008,colombo2009,krogstrup2013,holm2013}, solar water splitting cells\cite{wu2014}, light-emitting diodes\cite{Dimakis2014,herranz2020}, and lasers\cite{stettner2016,stettner2018,zhang2019}.

When NW shells are grown by directional deposition techniques such as molecular beam epitaxy (MBE), for which the previous references have been selected, material is deposited at a given moment only on those surface areas that are in direct line of sight for the impinging flux. In marked contrast, on a planar substrate the entire surface is covered simultaneously with the deposited material. It has been widely recognized that shadowing by neighboring NWs has to be taken into account for shell growth by directional deposition techniques, but otherwise little attention has been paid to this conceptual difference. As a notable exception,  shadowing by carefully designed bridge-like structures was lately utilized to provide a patterning of the superconductor shells deposited on semiconductor NWs, which improved the fabrication of hybrids for quantum transport studies.\cite{carrad2020} In a different approach to exploit the flux directionality of MBE for the synthesis of advanced NW structures, our group previously induced controlled NW bending by deliberate growth of a lattice-mismatched shell on only one NW sidewall\cite{lewis2018}. Typically, the substrate is rotated during growth, which for planar surfaces improves the thickness homogeneity. In the case of NW shell growth, rotation leads to subsequent exposure of all NW sidewalls to the incoming flux. For alloy growth, the constituent elements are provided by separate fluxes that impinge from different azimuthal directions. Consequently, the alloy constituents are \emph{sequentially} deposited on the NW sidewalls, as pointed out initially by Foxon \textit{et al.}\cite{Foxon2009} for the axial growth of GaN NWs. Recently, our group analyzed in detail how this sequentiality influences the morphology of GaN shells grown by MBE\cite{vantreeck2020}.

Here, we show for \ingaas/GaAs shells, a model candidate for an optically active quantum well (QW) shell\cite{Moewe2009,Dimakis2014,Park2014f,yan2015}, that the sequentiality of the As and the group-III fluxes in MBE has a crucial impact on the luminescence efficiency. By using two different As cells, we grow shells with different directions of the impinging fluxes and, thus, different deposition sequences under substrate rotation. The photoluminescence (PL) intensity of these samples differs by more than two orders of magnitude. We show that for the case leading to high luminescence intensity, the flux sequence resembles migration enhanced epitaxy (MEE), which has been used widely to achieve layers with highest material qualities even at low substrate temperatures\cite{KAWASHIMA1989a,Lopez1991,Hey2006}. 

\section*{Results and discussion}

\begin{figure}[t]
		\includegraphics[width=\columnwidth]{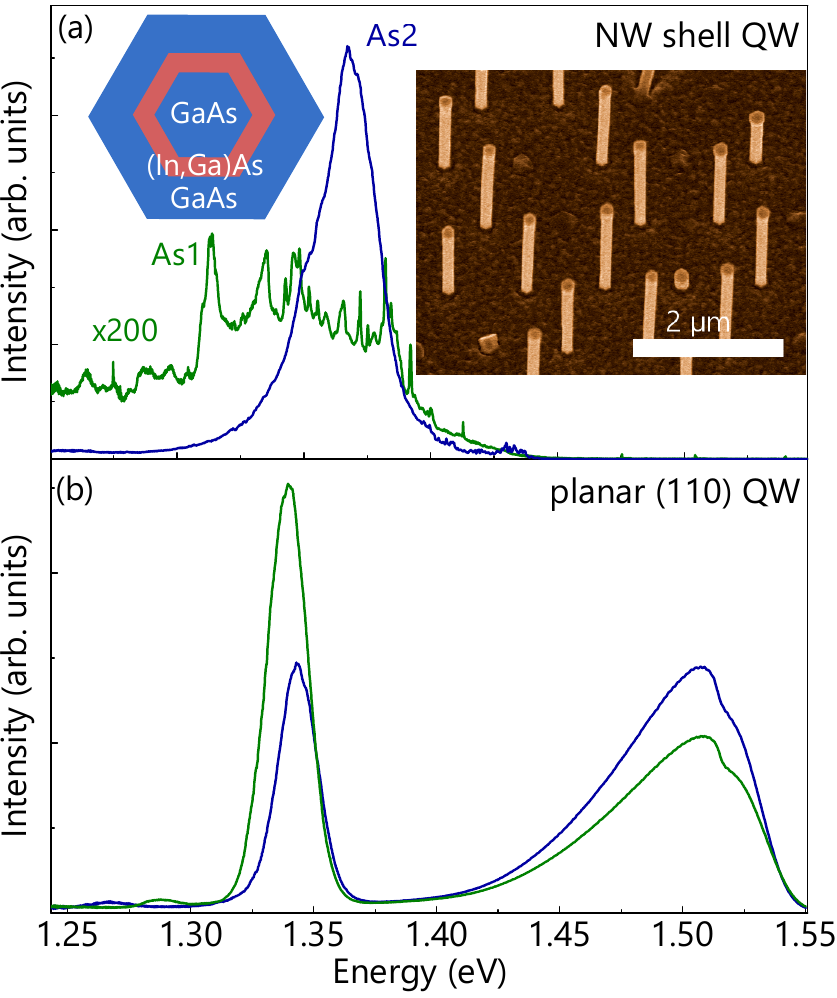}
		\caption{Comparison of PL spectra recorded at 10~K of NW and planar \inShell/GaAs QWs grown with different As cells: 
		(a) Core-shell NWs grown with either As source As1 (green curve) or As2 (blue curve). The PL intensity of the sample grown with the cell As1 has been multiplied with a factor of 200. The insets show the core-shell structure of the samples and a representative scanning electron micrograph of the NW arrays under investigation, tilted by 25$^{\circ}$ from the normal. 
		(b) Planar (110) QW samples with a structure equivalent to the shells of the NW samples grown with either As cell As1 (green curve) or As2 (blue curve).}
		\label{fig:Comparison}
\end{figure}	

The core-shell structure of the NWs under investigation is schematically depicted on the left-hand inset of Figure~\ref{fig:Comparison}a. A 10-nm thick \inShell shell QW is grown around a GaAs NW core with a diameter of 50~nm and covered with an outer GaAs shell of 30~nm thickness. The right-hand inset is a representative scanning electron micrograph of an ordered NW array with a nominal NW diameter and pitch of 130~nm and 1 \textmu m, respectively. It shows vertical NWs with a homogeneous thickness over their entire length and a bulky top, which is due to the crystallization of the Ga droplet after vapor-liquid-solid growth of the NW core. The NWs of the two samples that will be compared in the following exhibit a similar morphology, and we expect that the shells are coherent as demonstrated before\cite{Grandal2014}. Both samples were grown by MBE using the same growth conditions as detailed in the methods section, meaning equivalent NW core templates and the same substrate temperature and fluxes for the shell growth. The only difference between the two growth protocols is the choice of the As cell for the shell growth. 

\begin{figure}[t!]
		\includegraphics[width=\columnwidth]{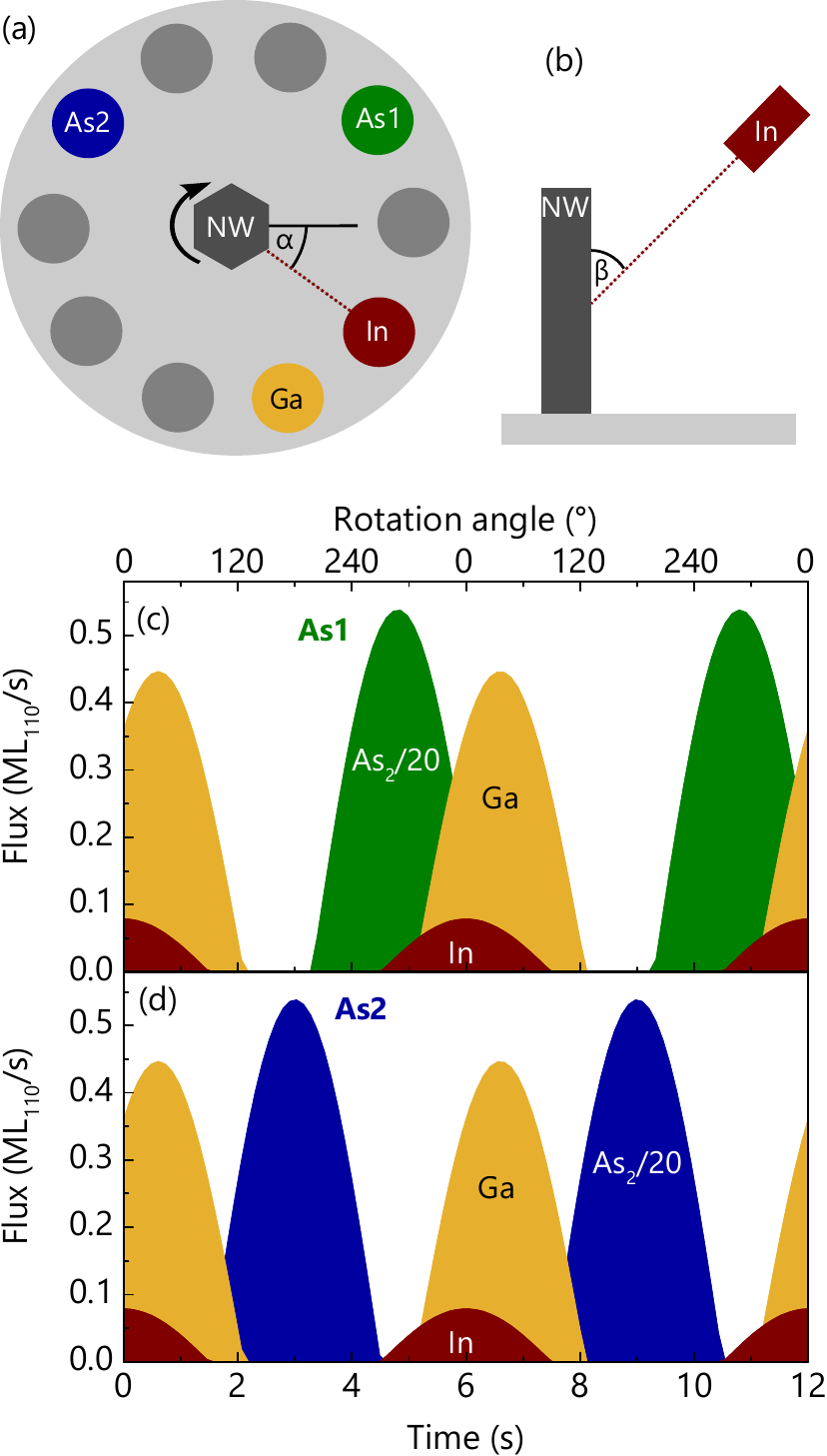}
		\caption{(a) Schematic of the cell configuration of the MBE system used for this study as seen from the top.
		(b) Schematic of the flux impinging on the NW sidewall as seen from the side.
		(c,d) Material fluxes impinging on a given NW side facet as a function of time and rotation angle as calculated by eq~\ref{eq:flux} using (c) cell As1 and (d) cell As2 for the experimental rotation speed of 6~rpm.
		}
		\label{fig:CellConfig}
\end{figure}

The PL spectra of these two samples presented in Figure~\ref{fig:Comparison}a show severe differences: The spectrum of the sample grown with cell As2 is dominated by an intense band at 1.37~eV associated with emission from the shell QW. In contrast, the PL intensity of the sample grown with source As1 is more than two orders of magnitude weaker. Furthermore, this spectrum consists of sharp lines in the spectral range 1.30~eV to 1.40~eV. We interpret these sharp lines as emission from excitons localized in the minima of potential fluctuations in the shell QW. Such localization is more pronounced in these \inShell/GaAs shell QWs than in conventional planar (100) \ingaas/GaAs QWs and occurs for low excitation density also in samples grown with cell As2, as we reported in a previous study\cite{kupers2019}. Hence, both the lower PL intensity and the broad band of sharp lines indicate a much lower photogenerated charge carrier concentration in the sample grown with cell As1. Since light coupling is identical for the two samples, the difference between the integrated PL intensities reveals directly that the internal quantum efficiency must be more than two orders of magnitude lower in the As1 sample. We emphasize that this difference is much larger than any variations resulting from changing the growth conditions when using a given As cell.

Figure~\ref{fig:Comparison}b depicts PL spectra of two planar reference samples containing an \inShell QW with the same thickness as the shell QW of the NW samples. In order to mimic the shell QWs as closely as possible, we used GaAs(110) substrates, which are of the same crystallographic orientation as the NW sidewall facets. The QWs of these two samples were grown using the same growth conditions as for the QWs in the NW samples, and the only difference between the two samples was again the choice of the As source. In stark contrast to the NW samples, these spectra are very similar and show an intense band at 1.34~eV with a full width at half maximum (FWHM) of 20~meV as well as a broader, weaker band around 1.50~eV. The first band can be assigned to the \inShell QW. The linewidth is larger compared to planar (100) QWs, which is commonly attributed to higher interface roughness resulting from growth in the [110] direction\cite{Someya1996,Hey2007}. The second broader band corresponds to luminescence from the n-type GaAs substrate. The similarity in the PL of the two planar samples rules out a possible contamination of one of the As sources as a reason for the drastic difference in PL intensity between the two NW samples.

The only difference between the two identical As sources is their location in the MBE system. Figure~\ref{fig:CellConfig}a shows a schematic of the locations of the relevant cells. The In and Ga cells are located next to each other. The As cell As1 is the next-nearest neighbour of the In cell, whereas the As cell As2 is placed exactly opposite to the In cell. The direction of the impinging fluxes relative to a given NW sidefacet can be characterized by two angles: 

The first, azimuthal angle $\alpha$ is described by the normal of the NW sidefacet and the direction of the effusion cell, as shown in Figure~\ref{fig:CellConfig}a. This angle varies continuously during substrate rotation. The second, polar angle $\beta$ is formed by the orientation of the cells with respect to the substrate normal, as sketched in Figure~\ref{fig:CellConfig}b. 

The flux impinging on one sidefacet is calculated as
\begin{strip}
\begin{align}
\label{eq:flux}
f_{\text{fac}}(t) = \begin{cases}
f_{\text{2D}}\tan(\beta)\cos[\alpha(t)-\alpha_i] &\text{for $-90^\circ < \alpha(t)-\alpha_i < 90^\circ $}\\
0 &\text{for $90^\circ \leq \alpha(t)-\alpha_i \leq 270^\circ $~,}
\end{cases}
\end{align}
\end{strip}
with the flux $f_{\text{2D}}$ impinging on the substrate plane and the cell-dependent angle $\alpha_i$ that accounts for the relative azimuthal cell positions. The fluxes impinging on one sidefacet calculated as a function of time are presented in Figure~\ref{fig:CellConfig}c for using cell As1 and in Figure~\ref{fig:CellConfig}d for using cell As2 as the As source, with a constant rotation speed $\dot{\alpha}=10~\text{rpm}=60^{\circ}/\text{s}$. In both sequences, the Ga and In fluxes overlap widely due to the cells' location next to each other. Using cell As1 [panel c], there is also a large overlap of the As flux with the In and Ga fluxes. In contrast, for cell As2 [panel d], there is no overlap of the As and In fluxes due to the opposite location of the cells to each other. Only for the Ga flux, there is a small overlap with the As flux. Consequently, using cell As1 all constituents are mostly co-deposited, whereas for using cell As2, As and In are never co-deposited.

The flux modulation with sequential group-III and As deposition (Figure~\ref{fig:CellConfig}d) resembles the one that is used for MEE of planar films. In that case, cell shutters are opened and closed such that the group-III and group-V fluxes impinge after each other on the substrate, with the aim to increase the diffusion of the group-III adatoms\cite{KAWASHIMA1989a}. We note that in comparison the effect of NW self-shadowing might be reduced to some extent by diffusion of group-III adatoms around the NW, but we know that this diffusion path is not necessarily very relevant since shells can be grown on only one side\cite{lewis2018}. The MEE approach is particularly beneficial for growth on GaAs(110) substrates, for which the As sticking factor is low, thus requiring low substrate temperatures and high V/III ratios to avoid surface facetting\cite{Allen1987,Wassermeier1994,Tok1997b}. Since these conditions limit the luminescence properties of the resulting films, MEE was used successfully for the growth of \ingaas \cite{Hey2006} layers on GaAs(110) with improved luminescence and compositional homogeneity compared to layers grown by co-deposition. At the same time, the improvement was not as drastic as the difference in PL observed here for NWs in Figure~\ref{fig:Comparison}a.

\begin{figure}[t]
		\centering
		\includegraphics[width=\columnwidth]{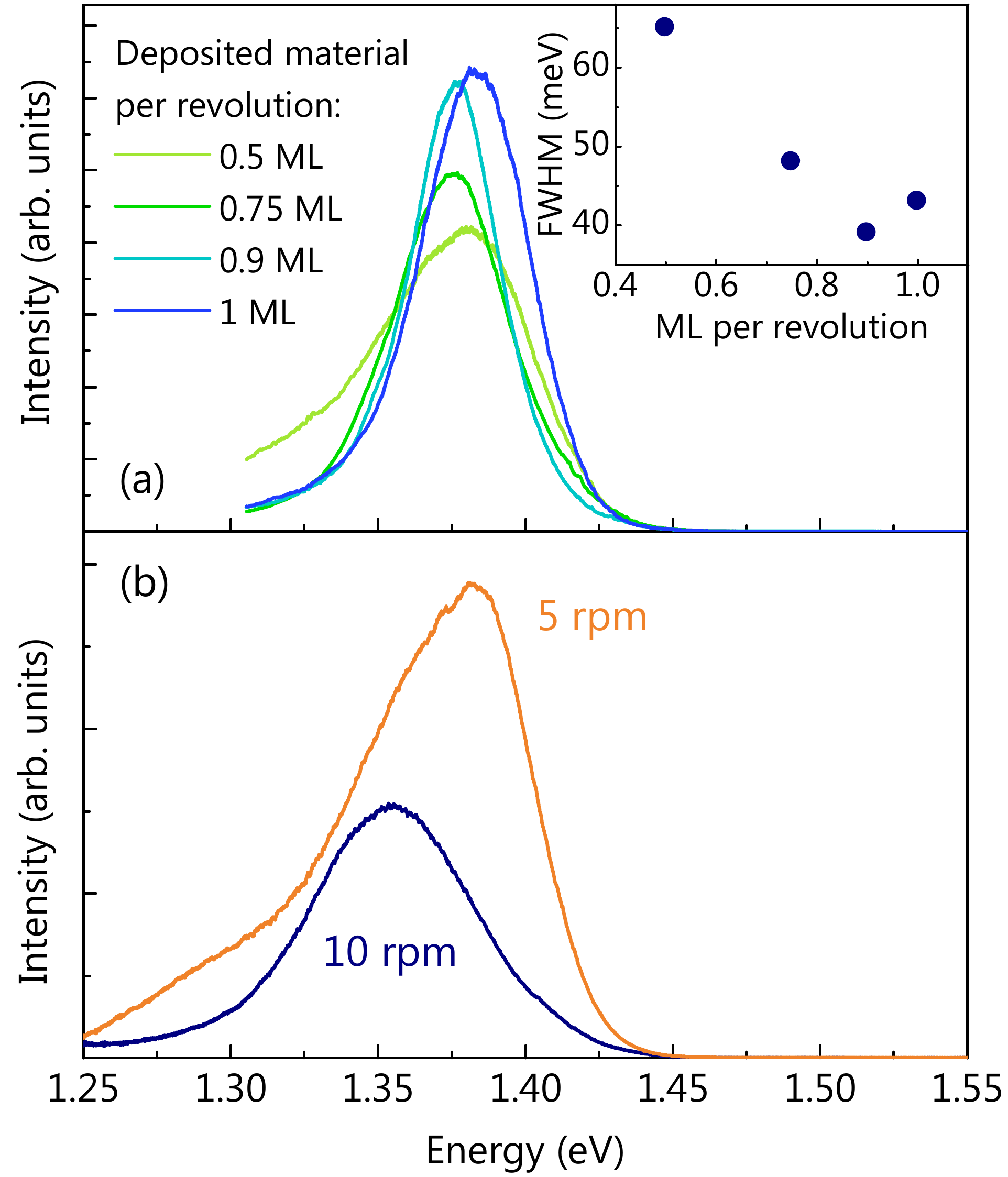}
		\caption{Analysis of other parameters affecting the sequentiality of deposition for shell growth with cell As2. 
		(a) Low-temperature (10~K) PL spectra of samples with shells grown at different growth rates, corresponding to different amounts of deposited material per revolution. Inset: FWHM of the spectra vs.\ the amount of material.
		(b)~Low-temperature (10~K) PL spectra taken on samples grown at varying rotation speed. We note that the growth conditions are slightly different compared to the samples shown in panel~(a), corresponding to 0.36 and 0.72~ML deposited per revolution.}
		\label{fig:GrowthRate}
\end{figure}	

The sequentiality of deposition sketched in Figure~\ref{fig:CellConfig}d is influenced in addition to the cell configuration by the growth rate and the rotation speed. Figure~\ref{fig:GrowthRate}a shows PL spectra taken on samples grown using cell As2 with different growth rates for the \inShell shell. For this sample series, the fluxes of all cells were adjusted to maintain the same V/III flux ratio. The varying growth rate corresponds to varying amounts of deposited material per revolution, given in the diagram in monolayers (MLs) in the $\langle 110 \rangle$ direction. All spectra show a single emission band at around 1.37~eV. The integrated intensity increases slightly with increasing growth rate and is highest for a full monolayer deposited per revolution. The inset in Figure~\ref{fig:GrowthRate}a displays the FWHM of these spectra, which decreases with increasing growth rate, but seems to be optimal for a deposition slightly below a full ML per revolution. The change in intensity is not significant, as it could be due to slight differences of the NW core template. However, the varying emission width implies that the \inShell shells are most homogeneous in terms of composition and/or interface roughness for the deposition of a complete layer per cycle. 

For the MEE growth of planar GaAs layers, it was shown that the recovery of the intensity in reflection high energy electron diffraction (RHEED) during growth was fastest for the deposition of complete Ga MLs per cycle\cite{Horikoshi1989}. At the same time, for the deposition of only 0.8 MLs per cycle, RHEED oscillations corresponding to the individual cycles were not continuously damped but regularly modulated with a longer period. These findings were interpreted to indicate that, even when the amount of material deposited per cycle does not correspond to a full ML, growth proceeds layer by layer, and the unevenness of the growth surface is at most one ML. Therefore, the variation in FWHM observed in our measurements is unlikely to be caused by increased interface roughness, and we ascribe the widening of the emission band rather to increased compositional inhomogeneity. 

Figure~\ref{fig:GrowthRate}b presents PL spectra taken on samples grown with different rotation speeds for the \inShell shell. We note that the substrate temperature and growth rate were slightly lower than for the previously discussed samples (see methods section). The sample grown with 10~rpm, as used for the previously discussed samples, exhibits a symmetric band at 1.35~eV. It is slightly red-shifted compared to the spectra in Figure~\ref{fig:GrowthRate}a, probably due to less In segregation at the lower substrate temperature\cite{Muraki1992}. The sample grown at the lower rotation speed of 5~rpm shows a peak at 1.38~eV with a higher intensity and a pronounced low-energy shoulder. By decreasing the rotation speed the amount of deposited material per rotation increases, here from 0.36 to 0.72~ML for the two samples. Thus, the sample with the larger amount of deposited material shows a more inhomogeneous emission, in contrast to the trend obtained for samples prepared with varying growth rate and constant rotation speed shown in Figure~\ref{fig:GrowthRate}a. Consequently, we cannot attribute the changes in emission observed in Figure~\ref{fig:GrowthRate}a to the change in deposited material per rotation. Therefore, we deduce that the change in sequence length itself has an important impact on the growth dynamics of the \inShell shell. 

During In and Ga deposition, the adatoms diffuse on the growth surface to reach energetically favourable locations. If this period is too long, In-rich clusters may form, giving rise to emission at low energies, as seen in the spectra. Likewise, for the analysis of GaN shell growth our group observed in simulations that increasing the rotation speed improves the homogeneity of the shell morphology\cite{vantreeck2020}. The experimental results presented now for \ingaas shells show that, even though the deposited amount per rotation and the rotation speed are connected, both parameters have a distinct impact on the growth dynamics.

\begin{figure}[t!]
		\includegraphics[width=\columnwidth]{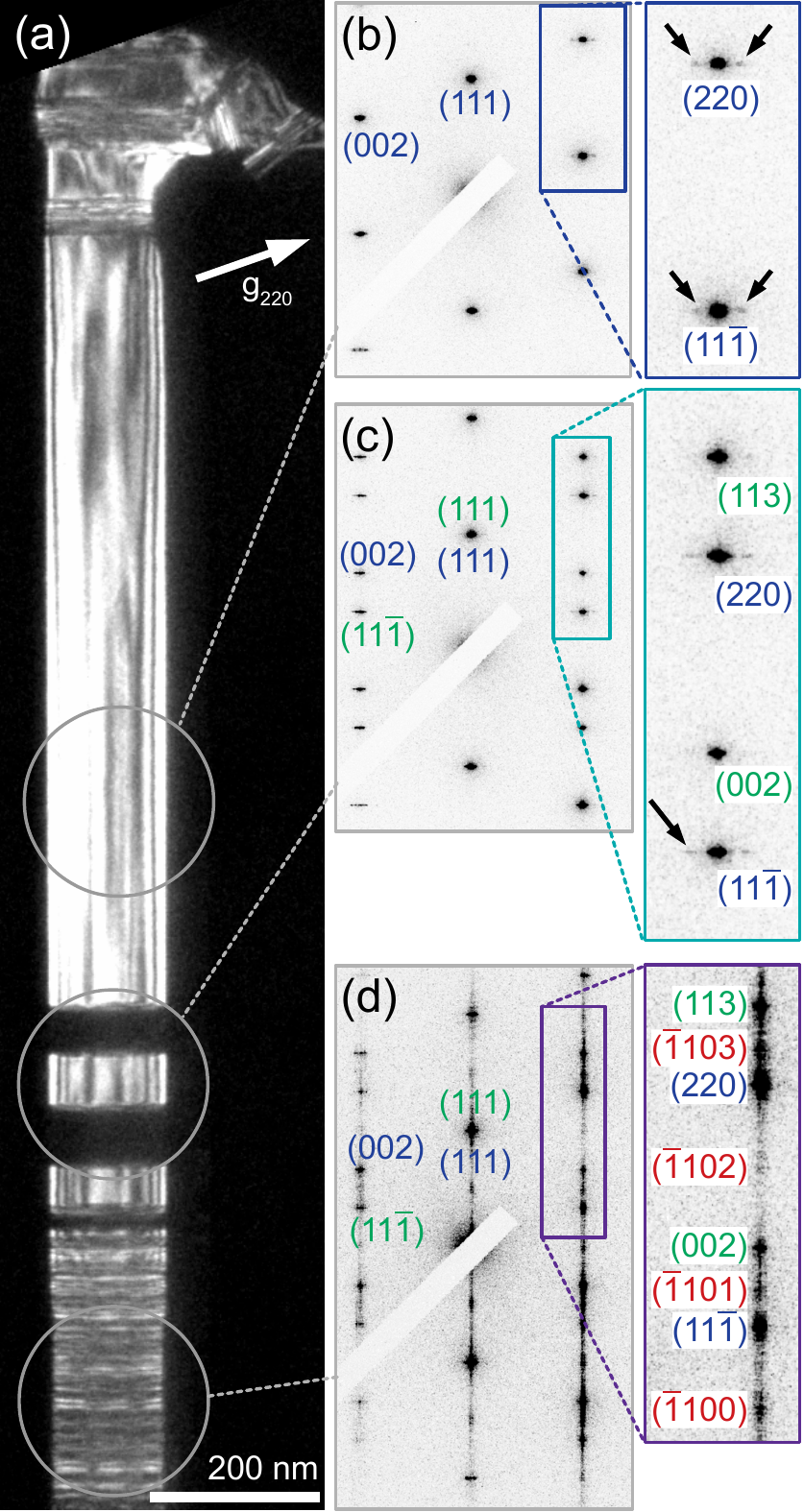}
		\caption[TEM measurement on a core-shell NW grown with cell As1]{(a) Transmission electron micrograph taken in dark-field mode on a core-shell NW. The three panels in (b), (c), and (d) show selected-area electron diffraction measurements at different positions of the NW. Here, the labels reveal the respective lattice vector in reciprocal space, and the arrows mark diffraction spots indicating a coherently strained shell.
		}
		\label{fig:TEM_shell}
\end{figure}

For planar \ingaas/GaAs(110) QWs, it has been reported that for an In content of 20\% plastic relaxation occurs in structures grown by conventional MBE, but is suppressed in MEE samples\cite{Hey2006}. In our case, the In content is lower, and, because of strain partitioning in the NW core-shell geometry, plastic relaxation is less likely. In order to investigate whether dislocations could be the reason for the low luminescence efficiency of the NWs grown with As cell As1, we analyzed such NWs by transmission electron microscopy (TEM). Figure~\ref{fig:TEM_shell}a shows a corresponding micrograph of a single, dispersed NW. The image is acquired under dark-field conditions using the cubic (220) diffraction spot. The NW top features an irregular widening, which we attribute to the consumption of the Ga droplet prior to shell growth.  At the bottom of the NW, axial contrast features with high frequency are detected. These features are attributed to the presence of stacking faults and thin slabs of different crystal phases. In contrast, the upper part of the NW does not show any clear axial contrast. Only the top of the NW exhibits deviations from a single crystal phase (like the bottom) due to the droplet consumption procedure. 

Figures \ref{fig:TEM_shell}b--d present selective area electron diffraction measurements that were acquired to identify the different crystal phases. For the measurement at the bottom shown in Figure~\ref{fig:TEM_shell}d, diffraction spots from both the twinned cubic zincblende structure (blue and green indices) and the hexagonal wurtzite structure (red indices) are detected, as well as streaks along the [111] direction, arising from the high density of twins and stacking faults. Figure~\ref{fig:TEM_shell}c depicts the measurement at the central part, exhibiting diffraction from twinned zincblende segments (blue and green). Figure~\ref{fig:TEM_shell}b presents the measurement at the top part of the NW where only the pure zincblende phase is detected (blue). The formation of different polytypes along the NW axis, as seen here, is often observed in NWs grown from droplets\cite{caroff2011}. Also, it is known that the shell adopts the crystal polytype of the respective core segment\cite{paladugu2009}.

The most important result of the TEM analysis is derived from the magnified diffraction reflections presented on the right-hand side of Figure~\ref{fig:TEM_shell}. There are secondary reflections to the right and left of the $(11\bar{1})$ and (220) zincblende reflections in panels (b) and (c). Their position exactly on a horizontal line with the main reflections corresponding to the radial NW direction indicates a coherent, elastically strained shell. Furthermore, dislocation lines would lead to contrast variations in the real-space image of panel (a). In the extended segments of the same polytype, the stripes along the NW axis correspond to thickness fringes due to the hexagonal shape of the NW, but there are not any indications of dislocations. We note, though, that for the bottom segment with mixed polytype dislocations are difficult to exclude on the basis of the present data. However, the lattice mismatch is essentially independent of polytype. Thus, the analysis of the extended phase-pure segments indicates that the low luminescence efficiency of the NWs grown with co-deposition, \textit{i.\,e.} cell As1, is not caused by extended defects.

\begin{figure*}
		\includegraphics[width=\textwidth]{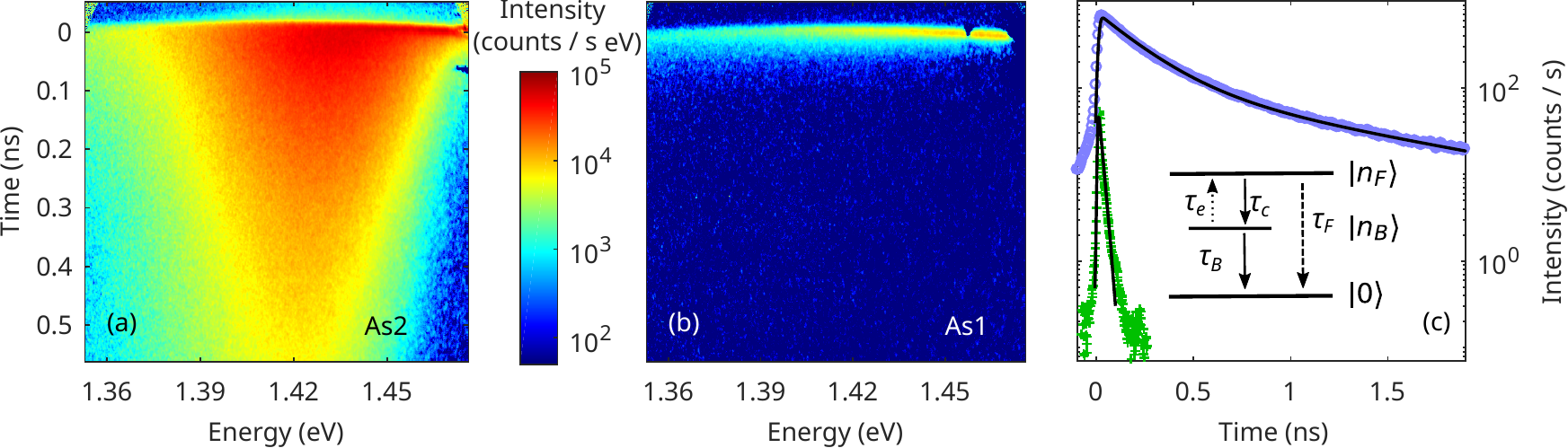}
		\caption{Analysis of the PL decay at 10~K for core-shell NWs grown with either cell As1 or As2. 
		(a) Streak camera image of the NWs grown with cell As2. 
		(b) Streak camera image of the NWs grown with cell As1.
 		(c) Transients of the PL intensity extracted from the data presented in (a) (blue symbols) and (b) (green symbols) by integrating across the entire spectral range. The lines are fits employing for the upper curve a coupled rate-equation system for free and bound excitons and for the lower curve a monoexponential decay. The inset depicts an energy level scheme visualizing the coupled free ($|n_F\rangle$) and bound ($|n_B\rangle$) exciton states involved in the radiative decay to the ground state $| 0 \rangle$. }
		\label{fig:TRPL}
\end{figure*}

A phenomenon known to limit the luminescence intensity of group-III arsenide NWs is nonradiative recombination at the large surface. However, in a detailed analysis we showed for GaAs NWs with \inShell shell QW that at 10~K an outer GaAs shell presents a sufficiently high barrier for charge carriers to suppress this channel\cite{kupers2019}. There are still two possible explanations for the drastic influence of the deposition sequentiality on the luminescence efficiency demonstrated in Figure~\ref{fig:Comparison}a. First, co-deposition as sketched in Figure~\ref{fig:CellConfig}c could result in a much higher density of deep point defects acting as non-radiative recombination centers, either throughout the QW shell or at its interfaces (or both). Second, this growth mode could lead to the incorporation of many more electrically active shallow point defects, most notably carbon. As a consequence, the charge carrier concentration in the NW would be increased, and, in combination with Fermi level pinning at the sidewalls, radial surface band bending would be much more pronounced. The associated internal electric field would reduce the overlap of electron and hole wavefunctions in the QW, thus decreasing the radiative recombination probability. These two explanations differ in how they reduce the internal quantum efficiency. In the first case, the nonradiative recombination rate is increased, in the second case, the radiative recombination rate is decreased. 

We carried out time-resolved PL measurements in order to identify the physical origin of the reduced luminescence efficiency for co-deposition. Figures~\ref{fig:TRPL}a,b display streak camera images obtained upon pulsed excitation at 10~K for the two core-shell NW arrays grown with cells As1 and As2. For both samples, the QW emission is detected at about 1.42~eV, \textit{i.\,e.}, at roughly 50~meV higher energy as compared to the emission obtained with continuous-wave excitation (cf.\ Figure~\ref{fig:Comparison}a). This blueshift is a result of the higher photoinduced carrier concentration for pulsed excitation and the entailing saturation of localized states.\cite{kupers2019} The most obvious difference between the samples is the time scale of the decay. To quantify this difference, we spectrally integrate the respective emission band and analyze the resulting PL transients as shown in Figure~\ref{fig:TRPL}c. In addition to the drastically different decay kinetics, which reflects the participation of nonradiative processes in the recombination, the peak intensity of the transients differs significantly. Since the two samples are basically identical in terms of NW diameter and pitch, the peak intensity is directly related to the radiative rate, \textit{i.\,e.}, the inverse radiative lifetime $\tau_{F,\text{r}}$ of the free exciton. 

For the sample grown with cell As2, we analyze the experimental transient with the coupled rate-equation system for free and bound excitons used previously for similar samples in Ref.~\citenum{kupers2019} and visualized in the inset of Fig.~\ref{fig:Comparison}c. The fit returns the time constants $\tau_c=0.56$, $\tau_e=1.18$, $\tau_F=0.34$, and $\tau_B=1.58$~ns. The time constants $\tau_c$ and $\tau_e$ characterizing the coupling of the two exciton states are essentially identical to those obtained in Ref.~\citenum{kupers2019}, whereas the effective lifetimes $\tau_F$ and $\tau_B$ of free and bound excitons, respectively, are a factor of two shorter. Taking the radiative lifetime $\tau_{F,\text{r}} =1.3$~ns of the free exciton as deduced in Ref.~\citenum{kupers2019} and assuming that the effective lifetime is lower due to an enhanced nonradiative contribution, we obtain a minimum nonradiative lifetime of $\tau_{F,\text{nr}} =0.46$~ns for the sample grown with cell As2. In comparison, the initial decay time found for the sample grown with cell As1 amounts to 10~ps. In addition, the peak intensity for the latter sample is reduced by an order of magnitude, \textit{i.\,e.}, the radiative lifetime exceeds 10~ns. Such an excessively long radiative lifetime for a free exciton in a QW clearly suggests the presence of substantial electric fields in the QW, decreasing the electron-hole wave function overlap and thus the radiative rate. Taken together, the integrated intensity for this sample is lower by more then two orders of magnitude due to the combined influence of the shorter nonradiative and the longer radiative lifetime. Our results thus suggest that the co-deposition of the alloy constituents during the growth of NW shells results in an increase of the densities of both shallow and deep point defects that act as dopants and nonradiative recombination centers, respectively. 

We note that the marked influence of the flux sequentiality inherent to NW shell growth on the luminescence efficiency is much more drastic than reported differences between planar layers grown by continuous MBE and MEE. Furthermore, for NW shells the impact of the flux sequentiality is stronger than that of a change in growth conditions for which only gradual variations were observed. Our findings suggest that the growth of NW shells is inherently different from planar growth on macroscopic substrates, even if the same crystallographic orientation is chosen. The biggest difference of NW sidewall facets compared to planar substrates are the six NW edges between the facets. These edges have a surface bonding configuration that differs from the (110) surface and that most likely leads to different growth kinetics. For the growth of \ingaas shells with high In content, it was shown that plastically relaxed clusters nucleate preferentially at these edges\cite{Lewis2017}, emphasizing their important role.

For the NWs discussed here, the edges always point radially in the $\left<11\bar{2}\right>$ directions. During shell growth, these edges may adopt the bonding configuration of the corresponding $\left\lbrace 11\bar{2} \right\rbrace $ surfaces, as it was reported for (In,Ga)As NWs\cite{Guo2013d} and (Al,Ga)As\cite{Heiss2013a}, (In,As)P\cite{Treu2013}, as well as InP shells \cite{Yuan2015b}. This surface is polar, so that a NW may exhibit three $\left\lbrace 112 \right\rbrace $A and three $\left\lbrace 112 \right\rbrace $B surfaces. For GaAs, it was shown that these surfaces are unstable due to their high surface energy and decompose into lower energy facets, either into a combination of $\left\lbrace 111 \right\rbrace $ and $\left\lbrace 113 \right\rbrace $ facets or into $\left\lbrace 110 \right\rbrace $ facets\cite{Geelhaar1999}. The choice of the facets depends on the crystal polarity and the chemical potential. The $(112)$A surface transforms into $\left\lbrace 111 \right\rbrace $ and $\left\lbrace 113 \right\rbrace $ facets under Ga-rich conditions and into $\left\lbrace 110 \right\rbrace $ facets under As-rich conditions. For the $(112)$B surface, the behaviour is the opposite. For $\left\lbrace 11\bar{2} \right\rbrace $ facets on NW edges during shell growth with co-deposition (using cell As1), the surface is always As-rich and, thus, depending on facet polarity, constantly either $\left\lbrace 110 \right\rbrace $ facets or a combination of $\left\lbrace 111 \right\rbrace $ and $\left\lbrace 113 \right\rbrace $ facets are energetically favoured. This constant preference for certain facets might result in the formation of nanofacets leading to rough surfaces and the formation of defects. In contrast, during shell growth with sequential deposition of group-III materials and As (using cell As2), the surface conditions change continuously from group-III-rich to As-rich. Under these conditions, the energetically preferred facets alternate with the flux sequence, possibly leading to a smoother surface. Hence, this dependence of the energetically favorable facet configuration on the flux sequence might explain the dramatic difference we observe in the luminescence efficiency. 

\section*{Conclusions}

The flux sequentiality inherent to the growth of NW shells by directional deposition techniques such as MBE has a drastic impact on the luminescence efficiency of \inShell/GaAs shell QWs. Only for the case that the group-III and As beams impinge from different directions, we obtained samples with bright luminescence. In this deposition geometry, the flux sequence on the NW sidefacets resembles MEE, which is known to improve for planar layers their smoothness and compositional homogeneity. The difference in luminescence efficiency of more than two orders of magnitude is caused by a higher density of both shallow and deep point defects acting as electrically active dopants and nonradiative recombination centers, respectively, that form for co-deposition. Our study shows that, for the growth of NW shells by MBE, the cell arrangement on the source flange is a decisive parameter, while it is only of technical relevance for the growth of planar layers. We emphasize that the dramatic effect on luminescence efficiency is not related to choices in growth protocols such as shutter sequences that are deliberately set for a given growth experiment, but is inherent to the self-shadowing of the NWs. Therefore, we expect that the source arrangement plays a role for the growth of NW shells in all alloy material systems, but the consequences may vary depending on the specific material properties. 

\section*{Methods}

Samples were grown in an MBE system containing ten cell ports, one with a Ga effusion cell and another one with an In effusion cell, and two occupied by valved cracker cells for supply of As$_2$. The As flux of both cells was calibrated by the transition of the surface reconstruction at the 1:1 point in GaAs(100) growth measured by RHEED. The substrate temperature was measured by optical pyrometry calibrated to the oxide desorption temperature of GaAs(100). More details on the calibration procedure can be found elsewhere\cite{Bastiman2016}. 

The NW samples were grown on Si(111) substrates covered by a patterned thermal oxide mask exhibiting hole patterns with 1~\textmu m separation. The GaAs NW cores were grown by Ga-assisted vapor-liquid-solid growth using a two-step procedure \cite{Kupers2018} leading to NWs with diameter of 50~nm and length of 2.5~$\upmu$m. After the core growth, the droplet was crystallized by only supplying As for 10~min at 630\celsius. Subsequently, the substrate temperature was decreased to 440\celsius under As flux exposure. For the growth of \inShell shells, In and Ga were supplied at fluxes corresponding to 15\% In, and the growth time was adjusted to achieve a nominal shell thickness of 10~nm. The As flux was varied to yield a V/III ratio of 20, which was optimal for optical emission properties of the \inShell shell. Finally, GaAs shells of 30~nm were grown subsequently around the \inShell shells, and the substrate temperature was reduced. For the samples grown with different rotation speeds presented in Figure~\ref{fig:GrowthRate}b, during shell growth the substrate temperature (420\celsius) and growth rate (0.2~ML/s) were slightly lower than for the other samples.

Planar reference samples were grown on n-type GaAs(110) substrates. First, a 100~nm thick buffer was grown, then a 10~nm thick \inShell layer was grown at a growth rate of 340~nm/h. Finally, a GaAs cap layer of 30~nm was grown. For all layers, the substrate temperature was 440\celsius, and the V/III ratio was 20. For all samples, the rotation speed during growth was 10~rotations per minute (rpm) if not noted otherwise.

Continous-wave micro-PL experiments were performed with samples mounted in a cold-finger cryostat, which was cooled to 10~K.  The He-Ne laser, emitting at 632.8~nm, was focused on the sample by a microscope objective with a numerical aperture of 0.25, resulting in the simultaneous excitation of about 25 NWs. The excitation power was varied between 66 and 710~\textmu W. The PL signal was collected by the same objective and dispersed by a monochromator. The signal was detected by a charge-coupled device and corrected for the system response of the setup. 

Time-resolved PL measurements were performed at 10~K by exciting the NW array with 650-nm pulses (200~fs width, 76~MHz repetition rate) from an optical parametric oscillator synchronously pumped by a Ti:sapphire laser, which itself is pumped by a frequency-doubled Nd:YVO$_{4}$ laser. The pulses with an energy of 130~pJ were focused onto the sample with a 50-mm plano-convex lens to a spot with a diameter of 40~\textmu m, resulting in the simultaneous excitation of about 300 NWs. The transient PL signal was dispersed by a monochromator and detected by a streak camera with a temporal resolution of either 5 or 50~ps.

TEM investigations were carried out at an acceleration voltage of 200~kV. For this purpose, NWs were scratched off from the substrate with tweezers and dispersed on a TEM sample grid with a thin carbon foil. Dark-field micrographs and selected-area diffraction patterns were acquired using a CCD camera.

\section*{Acknowledgments}

We are grateful to A.-K.~Bluhm for acquiring scanning electron microscopy images and to M.~Höricke and C.~Stemmler as well as A.~Wirsig for technical support at the MBE system. We thank O.~Krüger and M.~Matalla (Ferdinand-Braun-Institut, Berlin) for the substrate pre-patterning by electron beam lithography and A.~Taharoui as well as S.~Rauwerdink for substrate preparation. We appreciate the critical reading of the manuscript by M.~Ramsteiner. This study was partially funded by Deutsche Forschungsgemeinschaft under Grant Ge224/2, the Fonds National Suisse de la Rechereche Scientifique through Project No. 161032 (P.C.), and the Alexander von Humboldt Foundation (R.B.L.).

\bibliography{libMEE}

\providecommand{\latin}[1]{#1}
\makeatletter
\providecommand{\doi}
  {\begingroup\let\do\@makeother\dospecials
  \catcode`\{=1 \catcode`\}=2 \doi@aux}
\providecommand{\doi@aux}[1]{\endgroup\texttt{#1}}
\makeatother
\providecommand*\mcitethebibliography{\thebibliography}
\csname @ifundefined\endcsname{endmcitethebibliography}
  {\let\endmcitethebibliography\endthebibliography}{}
\begin{mcitethebibliography}{44}
\providecommand*\natexlab[1]{#1}
\providecommand*\mciteSetBstSublistMode[1]{}
\providecommand*\mciteSetBstMaxWidthForm[2]{}
\providecommand*\mciteBstWouldAddEndPuncttrue
  {\def\EndOfBibitem{\unskip.}}
\providecommand*\mciteBstWouldAddEndPunctfalse
  {\let\EndOfBibitem\relax}
\providecommand*\mciteSetBstMidEndSepPunct[3]{}
\providecommand*\mciteSetBstSublistLabelBeginEnd[3]{}
\providecommand*\EndOfBibitem{}
\mciteSetBstSublistMode{f}
\mciteSetBstMaxWidthForm{subitem}{(\alph{mcitesubitemcount})}
\mciteSetBstSublistLabelBeginEnd
  {\mcitemaxwidthsubitemform\space}
  {\relax}
  {\relax}

\bibitem[Garnett \latin{et~al.}(2019)Garnett, Mai, and Yang]{garnett2019}
Garnett,~E.; Mai,~L.; Yang,~P. Introduction: {{1D
  Nanomaterials}}/{{Nanowires}}. \emph{Chem. Rev.} \textbf{2019}, \emph{119},
  8955--8957\relax
\mciteBstWouldAddEndPuncttrue
\mciteSetBstMidEndSepPunct{\mcitedefaultmidpunct}
{\mcitedefaultendpunct}{\mcitedefaultseppunct}\relax
\EndOfBibitem
\bibitem[Royo \latin{et~al.}(2017)Royo, De~Luca, Rurali, and Zardo]{royo2017}
Royo,~M.; De~Luca,~M.; Rurali,~R.; Zardo,~I. A Review on
  {{III}}\textendash{{V}} Core\textendash Multishell Nanowires: Growth,
  Properties, and Applications. \emph{J. Phys. Appl. Phys.} \textbf{2017},
  \emph{50}, 143001\relax
\mciteBstWouldAddEndPuncttrue
\mciteSetBstMidEndSepPunct{\mcitedefaultmidpunct}
{\mcitedefaultendpunct}{\mcitedefaultseppunct}\relax
\EndOfBibitem
\bibitem[Waag \latin{et~al.}(2011)Waag, Wang, F{\"u}ndling, Ledig, Erenburg,
  Neumann, Al~Suleiman, Merzsch, Wei, Li, Wehmann, Bergbauer, Strassburg,
  Trampert, Jahn, and Riechert]{waag2011}
Waag,~A. \latin{et~al.}  The Nanorod Approach: {{GaN NanoLEDs}} for Solid State
  Lighting. \emph{Phys. Status Solidi C} \textbf{2011}, \emph{8},
  2296--2301\relax
\mciteBstWouldAddEndPuncttrue
\mciteSetBstMidEndSepPunct{\mcitedefaultmidpunct}
{\mcitedefaultendpunct}{\mcitedefaultseppunct}\relax
\EndOfBibitem
\bibitem[Kavanagh(2010)]{Kavanagh2010}
Kavanagh,~K.~L. Misfit Dislocations in Nanowire Heterostructures.
  \emph{Semicond. Sci. Technol.} \textbf{2010}, \emph{25}, 024006\relax
\mciteBstWouldAddEndPuncttrue
\mciteSetBstMidEndSepPunct{\mcitedefaultmidpunct}
{\mcitedefaultendpunct}{\mcitedefaultseppunct}\relax
\EndOfBibitem
\bibitem[Czaban \latin{et~al.}(2008)Czaban, Thompson, and LaPierre]{czaban2008}
Czaban,~J.~A.; Thompson,~D.~A.; LaPierre,~R.~R. {{GaAs}} Core- Shell Nanowires
  for Photovoltaic Applications. \emph{Nano Lett.} \textbf{2008}, \emph{9},
  148--154\relax
\mciteBstWouldAddEndPuncttrue
\mciteSetBstMidEndSepPunct{\mcitedefaultmidpunct}
{\mcitedefaultendpunct}{\mcitedefaultseppunct}\relax
\EndOfBibitem
\bibitem[Colombo \latin{et~al.}(2009)Colombo, Hei{\ss}, Gr{\"a}tzel,
  {Fontcuberta i Morral}, and Gr{\"a}tzel]{colombo2009}
Colombo,~C.; Hei{\ss},~M.; Gr{\"a}tzel,~M.; {Fontcuberta i Morral},~A.;
  Gr{\"a}tzel,~M. Gallium Arsenide P-i-n Radial Structures for Photovoltaic
  Applications. \emph{Appl. Phys. Lett.} \textbf{2009}, \emph{94}, 173108\relax
\mciteBstWouldAddEndPuncttrue
\mciteSetBstMidEndSepPunct{\mcitedefaultmidpunct}
{\mcitedefaultendpunct}{\mcitedefaultseppunct}\relax
\EndOfBibitem
\bibitem[Krogstrup \latin{et~al.}(2013)Krogstrup, J{\o}rgensen, Heiss,
  Demichel, Holm, Aagesen, Nygard, and {Fontcuberta i Morral}]{krogstrup2013}
Krogstrup,~P.; J{\o}rgensen,~H.~I.; Heiss,~M.; Demichel,~O.; Holm,~J.~V.;
  Aagesen,~M.; Nygard,~J.; {Fontcuberta i Morral},~A. Single-Nanowire Solar
  Cells beyond the {{Shockley}}\textendash{{Queisser}} Limit. \emph{Nat.
  Photonics} \textbf{2013}, \emph{7}, 306--310\relax
\mciteBstWouldAddEndPuncttrue
\mciteSetBstMidEndSepPunct{\mcitedefaultmidpunct}
{\mcitedefaultendpunct}{\mcitedefaultseppunct}\relax
\EndOfBibitem
\bibitem[Holm \latin{et~al.}(2013)Holm, J{\o}rgensen, Krogstrup, Nyg{\aa}rd,
  Liu, and Aagesen]{holm2013}
Holm,~J.~V.; J{\o}rgensen,~H.~I.; Krogstrup,~P.; Nyg{\aa}rd,~J.; Liu,~H.;
  Aagesen,~M. Surface-Passivated {{GaAsP}} Single-Nanowire Solar Cells
  Exceeding 10\% Efficiency Grown on Silicon. \emph{Nat. Commun.}
  \textbf{2013}, \emph{4}, 1498\relax
\mciteBstWouldAddEndPuncttrue
\mciteSetBstMidEndSepPunct{\mcitedefaultmidpunct}
{\mcitedefaultendpunct}{\mcitedefaultseppunct}\relax
\EndOfBibitem
\bibitem[Wu \latin{et~al.}(2014)Wu, Li, Kubota, Domen, Aagesen, Ward, Sanchez,
  Beanland, Zhang, Tang, Hatch, Seeds, and Liu]{wu2014}
Wu,~J.; Li,~Y.; Kubota,~J.; Domen,~K.; Aagesen,~M.; Ward,~T.; Sanchez,~A.;
  Beanland,~R.; Zhang,~Y.; Tang,~M.; Hatch,~S.; Seeds,~A.; Liu,~H. Wafer-Scale
  Fabrication of Self-Catalyzed 1.7 {{eV GaAsP}} Core-Shell Nanowire
  Photocathode on Silicon Substrates. \emph{Nano Lett.} \textbf{2014},
  \emph{14}, 2013--8\relax
\mciteBstWouldAddEndPuncttrue
\mciteSetBstMidEndSepPunct{\mcitedefaultmidpunct}
{\mcitedefaultendpunct}{\mcitedefaultseppunct}\relax
\EndOfBibitem
\bibitem[Dimakis \latin{et~al.}(2014)Dimakis, Jahn, Ramsteiner, Tahraoui,
  Grandal, Kong, Marquardt, Trampert, Riechert, and Geelhaar]{Dimakis2014}
Dimakis,~E.; Jahn,~U.; Ramsteiner,~M.; Tahraoui,~A.; Grandal,~J.; Kong,~X.;
  Marquardt,~O.; Trampert,~A.; Riechert,~H.; Geelhaar,~L. Coaxial Multishell
  ({{In}},{{Ga}}){{As}}/{{GaAs}} Nanowires for near-Infrared Emission on Si
  Substrates. \emph{Nano Lett.} \textbf{2014}, \emph{14}, 2604--2609\relax
\mciteBstWouldAddEndPuncttrue
\mciteSetBstMidEndSepPunct{\mcitedefaultmidpunct}
{\mcitedefaultendpunct}{\mcitedefaultseppunct}\relax
\EndOfBibitem
\bibitem[Herranz \latin{et~al.}(2020)Herranz, Corfdir, Luna, Jahn, Lewis,
  Schrottke, L{\"a}hnemann, Tahraoui, Trampert, Brandt, and
  Geelhaar]{herranz2020}
Herranz,~J.; Corfdir,~P.; Luna,~E.; Jahn,~U.; Lewis,~R.~B.; Schrottke,~L.;
  L{\"a}hnemann,~J.; Tahraoui,~A.; Trampert,~A.; Brandt,~O.; Geelhaar,~L.
  Coaxial {{GaAs}}/({{In}},{{Ga}}){{As}} Dot-in-a-Well Nanowire
  Heterostructures for Electrically Driven Infrared Light Generation on {{Si}}
  in the Telecommunication {{O Band}}. \emph{ACS Appl. Nano Mater.}
  \textbf{2020}, \emph{3}, 165--174\relax
\mciteBstWouldAddEndPuncttrue
\mciteSetBstMidEndSepPunct{\mcitedefaultmidpunct}
{\mcitedefaultendpunct}{\mcitedefaultseppunct}\relax
\EndOfBibitem
\bibitem[Stettner \latin{et~al.}(2016)Stettner, Zimmermann, Loitsch,
  D{\"o}blinger, Regler, Mayer, Winnerl, Matich, Riedl, Kaniber, Abstreiter,
  Koblm{\"u}ller, and Finley]{stettner2016}
Stettner,~T.; Zimmermann,~P.; Loitsch,~B.; D{\"o}blinger,~M.; Regler,~A.;
  Mayer,~B.; Winnerl,~J.; Matich,~S.; Riedl,~H.; Kaniber,~M.; Abstreiter,~G.;
  Koblm{\"u}ller,~G.; Finley,~J.~J. Coaxial {{GaAs}}-{{AlGaAs}} Core-Multishell
  Nanowire Lasers with Epitaxial Gain Control. \emph{Appl. Phys. Lett.}
  \textbf{2016}, \emph{108}, 011108\relax
\mciteBstWouldAddEndPuncttrue
\mciteSetBstMidEndSepPunct{\mcitedefaultmidpunct}
{\mcitedefaultendpunct}{\mcitedefaultseppunct}\relax
\EndOfBibitem
\bibitem[Stettner \latin{et~al.}(2018)Stettner, Thurn, D{\"o}blinger, Hill,
  Bissinger, Schmiedeke, Matich, Kostenbader, Ruhstorfer, Riedl, Kaniber,
  Lauhon, Finley, and Koblm{\"u}ller]{stettner2018}
Stettner,~T.; Thurn,~A.; D{\"o}blinger,~M.; Hill,~M.~O.; Bissinger,~J.;
  Schmiedeke,~P.; Matich,~S.; Kostenbader,~T.; Ruhstorfer,~D.; Riedl,~H.;
  Kaniber,~M.; Lauhon,~L.~J.; Finley,~J.~J.; Koblm{\"u}ller,~G. Tuning Lasing
  Emission toward Long Wavelengths in {{GaAs}}-({{In}},{{Al}}){{GaAs}}
  Core-Multishell Nanowires. \emph{Nano Lett.} \textbf{2018}, \emph{18},
  6292--6300\relax
\mciteBstWouldAddEndPuncttrue
\mciteSetBstMidEndSepPunct{\mcitedefaultmidpunct}
{\mcitedefaultendpunct}{\mcitedefaultseppunct}\relax
\EndOfBibitem
\bibitem[Zhang \latin{et~al.}(2019)Zhang, Davis, Fonseka, Velichko, Gustafsson,
  Godde, Saxena, Aagesen, Parkinson, Gott, Huo, Sanchez, Mowbray, and
  Liu]{zhang2019}
Zhang,~Y.; Davis,~G.; Fonseka,~H.~A.; Velichko,~A.; Gustafsson,~A.; Godde,~T.;
  Saxena,~D.; Aagesen,~M.; Parkinson,~P.~W.; Gott,~J.~A.; Huo,~S.;
  Sanchez,~A.~M.; Mowbray,~D.~J.; Liu,~H. Highly {{Strained
  III}}\textendash{{V}}\textendash{{V Coaxial Nanowire Quantum Wells}} with
  {{Strong Carrier Confinement}}. \emph{ACS Nano} \textbf{2019}, \emph{13},
  5931--5938\relax
\mciteBstWouldAddEndPuncttrue
\mciteSetBstMidEndSepPunct{\mcitedefaultmidpunct}
{\mcitedefaultendpunct}{\mcitedefaultseppunct}\relax
\EndOfBibitem
\bibitem[Carrad \latin{et~al.}(2020)Carrad, Bjergfelt, Kanne, Aagesen, Krizek,
  Fiordaliso, Johnson, Nyg{\aa}rd, and Jespersen]{carrad2020}
Carrad,~D.~J.; Bjergfelt,~M.; Kanne,~T.; Aagesen,~M.; Krizek,~F.;
  Fiordaliso,~E.~M.; Johnson,~E.; Nyg{\aa}rd,~J.; Jespersen,~T.~S. Shadow
  {{Epitaxy}} for {{In Situ Growth}} of {{Generic
  Semiconductor}}/{{Superconductor Hybrids}}. \emph{Adv. Mater.} \textbf{2020},
  1908411\relax
\mciteBstWouldAddEndPuncttrue
\mciteSetBstMidEndSepPunct{\mcitedefaultmidpunct}
{\mcitedefaultendpunct}{\mcitedefaultseppunct}\relax
\EndOfBibitem
\bibitem[Lewis \latin{et~al.}(2018)Lewis, Corfdir, K{\"u}pers, Flissikowski,
  Brandt, and Geelhaar]{lewis2018}
Lewis,~R.~B.; Corfdir,~P.; K{\"u}pers,~H.; Flissikowski,~T.; Brandt,~O.;
  Geelhaar,~L. Nanowires Bending over Backward from Strain Partitioning in
  Asymmetric Core-Shell Heterostructures. \emph{Nano Lett.} \textbf{2018},
  \emph{18}, 2343--2350\relax
\mciteBstWouldAddEndPuncttrue
\mciteSetBstMidEndSepPunct{\mcitedefaultmidpunct}
{\mcitedefaultendpunct}{\mcitedefaultseppunct}\relax
\EndOfBibitem
\bibitem[Foxon \latin{et~al.}(2009)Foxon, Novikov, Hall, Campion, Cherns,
  Griffiths, and Khongphetsak]{Foxon2009}
Foxon,~C.~T.; Novikov,~S.; Hall,~J.; Campion,~R.; Cherns,~D.; Griffiths,~I.;
  Khongphetsak,~S. A Complementary Geometric Model for the Growth of {{GaN}}
  Nanocolumns Prepared by Plasma-Assisted Molecular Beam Epitaxy. \emph{J.
  Cryst. Growth} \textbf{2009}, \emph{311}, 3423--3427\relax
\mciteBstWouldAddEndPuncttrue
\mciteSetBstMidEndSepPunct{\mcitedefaultmidpunct}
{\mcitedefaultendpunct}{\mcitedefaultseppunct}\relax
\EndOfBibitem
\bibitem[{van Treeck} \latin{et~al.}(2020){van Treeck},
  {Fern{\'a}ndez-Garrido}, and Geelhaar]{vantreeck2020}
{van Treeck},~D.; {Fern{\'a}ndez-Garrido},~S.; Geelhaar,~L. Influence of the
  Source Arrangement on Shell Growth around {{GaN}} Nanowires in Molecular Beam
  Epitaxy. \emph{Phys. Rev. Materials} \textbf{2020}, \emph{4}, 013404\relax
\mciteBstWouldAddEndPuncttrue
\mciteSetBstMidEndSepPunct{\mcitedefaultmidpunct}
{\mcitedefaultendpunct}{\mcitedefaultseppunct}\relax
\EndOfBibitem
\bibitem[Moewe \latin{et~al.}(2009)Moewe, Chuang, Crankshaw, Ng, and
  {Chang-Hasnain}]{Moewe2009}
Moewe,~M.; Chuang,~L.~C.; Crankshaw,~S.; Ng,~K.~W.; {Chang-Hasnain},~C.
  Core-Shell {{InGaAs}}/{{GaAs}} Quantum Well Nanoneedles Grown on Silicon with
  Silicon-Transparent Emission. \emph{Opt. Express} \textbf{2009}, \emph{17},
  7831\relax
\mciteBstWouldAddEndPuncttrue
\mciteSetBstMidEndSepPunct{\mcitedefaultmidpunct}
{\mcitedefaultendpunct}{\mcitedefaultseppunct}\relax
\EndOfBibitem
\bibitem[Park \latin{et~al.}(2014)Park, Park, Ravindran, Jang, Jo, Kim, and
  Lee]{Park2014f}
Park,~K.~W.; Park,~C.~Y.; Ravindran,~S.; Jang,~J.-S.; Jo,~Y.-R.; Kim,~B.-J.;
  Lee,~Y.~T. Observation and Tunability of Room Temperature Photoluminescence
  of {{GaAs}}/{{GaInAs}} Core-Multiple-Quantum-Well Shell Nanowire Structure
  Grown on {{Si}} (100) by Molecular Beam Epitaxy. \emph{Nanoscale Res. Lett.}
  \textbf{2014}, \emph{9}, 626\relax
\mciteBstWouldAddEndPuncttrue
\mciteSetBstMidEndSepPunct{\mcitedefaultmidpunct}
{\mcitedefaultendpunct}{\mcitedefaultseppunct}\relax
\EndOfBibitem
\bibitem[Yan \latin{et~al.}(2015)Yan, Zhang, Li, Wu, Cui, and Ren]{yan2015}
Yan,~X.; Zhang,~X.; Li,~J.; Wu,~Y.; Cui,~J.; Ren,~X. Fabrication and Optical
  Properties of {{GaAs}}/{{InGaAs}}/{{GaAs}} Nanowire Core-Multishell Quantum
  Well Heterostructures. \emph{Nanoscale} \textbf{2015}, \emph{7},
  1110--5\relax
\mciteBstWouldAddEndPuncttrue
\mciteSetBstMidEndSepPunct{\mcitedefaultmidpunct}
{\mcitedefaultendpunct}{\mcitedefaultseppunct}\relax
\EndOfBibitem
\bibitem[Horikoshi \latin{et~al.}(1988)Horikoshi, Kawashima, and
  Yamaguchi]{KAWASHIMA1989a}
Horikoshi,~Y.; Kawashima,~M.; Yamaguchi,~H. Migration-Enhanced Epitaxy of
  {{GaAs}} and {{AlGaAs}}. \emph{Jpn. J. Appl. Phys.} \textbf{1988}, \emph{27},
  169--179\relax
\mciteBstWouldAddEndPuncttrue
\mciteSetBstMidEndSepPunct{\mcitedefaultmidpunct}
{\mcitedefaultendpunct}{\mcitedefaultseppunct}\relax
\EndOfBibitem
\bibitem[L{\'o}pez \latin{et~al.}(1991)L{\'o}pez, Takano, Pak, and
  Yonezu]{Lopez1991}
L{\'o}pez,~M.; Takano,~Y.; Pak,~K.; Yonezu,~H. Realization of Low Facet Density
  and the Growth Mechanism of {{GaAs}} on {{GaAs}}(110) by Migration-Enhanced
  Epitaxy. \emph{Appl. Phys. Lett.} \textbf{1991}, \emph{58}, 580--582\relax
\mciteBstWouldAddEndPuncttrue
\mciteSetBstMidEndSepPunct{\mcitedefaultmidpunct}
{\mcitedefaultendpunct}{\mcitedefaultseppunct}\relax
\EndOfBibitem
\bibitem[Hey \latin{et~al.}(2006)Hey, Trampert, and Santos]{Hey2006}
Hey,~R.; Trampert,~a.; Santos,~P. ({{In}},{{Ga}}){{As}}/{{GaAs}} Quantum Wells
  on {{GaAs}}(110). \emph{Phys. Status Solidi C} \textbf{2006}, \emph{3},
  651--654\relax
\mciteBstWouldAddEndPuncttrue
\mciteSetBstMidEndSepPunct{\mcitedefaultmidpunct}
{\mcitedefaultendpunct}{\mcitedefaultseppunct}\relax
\EndOfBibitem
\bibitem[Grandal \latin{et~al.}(2014)Grandal, Wu, Kong, Hanke, Dimakis,
  Geelhaar, Riechert, and Trampert]{Grandal2014}
Grandal,~J.; Wu,~M.; Kong,~X.; Hanke,~M.; Dimakis,~E.; Geelhaar,~L.;
  Riechert,~H.; Trampert,~A. Plan-View Transmission Electron Microscopy
  Investigation of {{GaAs}}/({{In}},{{Ga}}){{As}} Core-Shell Nanowires.
  \emph{Appl. Phys. Lett.} \textbf{2014}, \emph{105}, 121602\relax
\mciteBstWouldAddEndPuncttrue
\mciteSetBstMidEndSepPunct{\mcitedefaultmidpunct}
{\mcitedefaultendpunct}{\mcitedefaultseppunct}\relax
\EndOfBibitem
\bibitem[K{\"u}pers \latin{et~al.}(2019)K{\"u}pers, Corfdir, Lewis,
  Flissikowski, Tahraoui, Grahn, Brandt, and Geelhaar]{kupers2019}
K{\"u}pers,~H.; Corfdir,~P.; Lewis,~R.~B.; Flissikowski,~T.; Tahraoui,~A.;
  Grahn,~H.~T.; Brandt,~O.; Geelhaar,~L. Impact of Outer Shell Structure and
  Localization Effects on Charge Carrier Dynamics in
  {{GaAs}}/({{In}},{{Ga}}){{As}} Nanowire Core\textendash Shell Quantum Wells.
  \emph{physica status solidi (RRL)} \textbf{2019}, \emph{13}, 1800527\relax
\mciteBstWouldAddEndPuncttrue
\mciteSetBstMidEndSepPunct{\mcitedefaultmidpunct}
{\mcitedefaultendpunct}{\mcitedefaultseppunct}\relax
\EndOfBibitem
\bibitem[Someya \latin{et~al.}(1996)Someya, Akiyama, and Sakaki]{Someya1996}
Someya,~T.; Akiyama,~H.; Sakaki,~H. Molecular Beam Epitaxial Growth of in 0.15
  Ga 0.85 as Quantum Wells on (110) {{GaAs}} Surfaces. \emph{Jpn. J. Appl.
  Phys.} \textbf{1996}, \emph{35}, 2544--2547\relax
\mciteBstWouldAddEndPuncttrue
\mciteSetBstMidEndSepPunct{\mcitedefaultmidpunct}
{\mcitedefaultendpunct}{\mcitedefaultseppunct}\relax
\EndOfBibitem
\bibitem[Hey \latin{et~al.}(2007)Hey, Trampert, Jahn, Couto, and
  Santos]{Hey2007}
Hey,~R.; Trampert,~A.; Jahn,~U.; Couto,~O.; Santos,~P. Growth of
  ({{In}},{{Ga}}){{As}}/({{Al}},{{Ga}}){{As}} Quantum Wells on {{GaAs}}(110) by
  {{MBE}}. \emph{J. Cryst. Growth} \textbf{2007}, \emph{301-302},
  158--162\relax
\mciteBstWouldAddEndPuncttrue
\mciteSetBstMidEndSepPunct{\mcitedefaultmidpunct}
{\mcitedefaultendpunct}{\mcitedefaultseppunct}\relax
\EndOfBibitem
\bibitem[Allen \latin{et~al.}(1987)Allen, Weber, Washburn, and Pao]{Allen1987}
Allen,~L. T.~P.; Weber,~E.~R.; Washburn,~J.; Pao,~Y.~C. Device Quality Growth
  and Characterization of (110) {{GaAs}} Grown by Molecular Beam Epitaxy.
  \emph{Appl. Phys. Lett.} \textbf{1987}, \emph{51}, 670--672\relax
\mciteBstWouldAddEndPuncttrue
\mciteSetBstMidEndSepPunct{\mcitedefaultmidpunct}
{\mcitedefaultendpunct}{\mcitedefaultseppunct}\relax
\EndOfBibitem
\bibitem[Wassermeier \latin{et~al.}(1994)Wassermeier, Yang, Tourni{\'e},
  D{\"a}weritz, and Ploog]{Wassermeier1994}
Wassermeier,~M.; Yang,~H.; Tourni{\'e},~E.; D{\"a}weritz,~L.; Ploog,~K. Growth
  Mechanism of {{GaAs}} on (110) {{GaAs}} Studied by High-Energy Electron
  Diffraction and Atomic Force Microscopy. \emph{J. Vac. Sci. Technol. B
  Microelectron. Nanometer Struct. Process. Meas. Phenom.} \textbf{1994},
  \emph{12}, 2574--2578\relax
\mciteBstWouldAddEndPuncttrue
\mciteSetBstMidEndSepPunct{\mcitedefaultmidpunct}
{\mcitedefaultendpunct}{\mcitedefaultseppunct}\relax
\EndOfBibitem
\bibitem[Tok \latin{et~al.}(1997)Tok, Jones, Neave, Zhang, and
  a.~Joyce]{Tok1997b}
Tok,~E.~S.; Jones,~T.~S.; Neave,~J.~H.; Zhang,~J.; a.~Joyce,~B. Is the Arsenic
  Incorporation Kinetics Important When Growing {{GaAs}}(001), (110), and
  (111){{A}} Films? \emph{Appl. Phys. Lett.} \textbf{1997}, \emph{71},
  3278\relax
\mciteBstWouldAddEndPuncttrue
\mciteSetBstMidEndSepPunct{\mcitedefaultmidpunct}
{\mcitedefaultendpunct}{\mcitedefaultseppunct}\relax
\EndOfBibitem
\bibitem[Horikoshi(1993)]{Horikoshi1989}
Horikoshi,~Y. Migration-Enhanced Epitaxy of {{GaAs}} and {{AlGaAs}}.
  \emph{Semicond. Sci. Technol.} \textbf{1993}, \emph{8}, 1032--1051\relax
\mciteBstWouldAddEndPuncttrue
\mciteSetBstMidEndSepPunct{\mcitedefaultmidpunct}
{\mcitedefaultendpunct}{\mcitedefaultseppunct}\relax
\EndOfBibitem
\bibitem[Muraki \latin{et~al.}(1992)Muraki, Fukatsu, Shiraki, and
  Ito]{Muraki1992}
Muraki,~K.; Fukatsu,~S.; Shiraki,~Y.; Ito,~R. Surface Segregation of {{In}}
  Atoms during Molecular Beam Epitaxy and Its Influence on the Energy Levels in
  {{InGaAs}}/{{GaAs}} Quantum Wells. \emph{Appl. Phys. Lett.} \textbf{1992},
  \emph{61}, 557\relax
\mciteBstWouldAddEndPuncttrue
\mciteSetBstMidEndSepPunct{\mcitedefaultmidpunct}
{\mcitedefaultendpunct}{\mcitedefaultseppunct}\relax
\EndOfBibitem
\bibitem[Caroff \latin{et~al.}(2011)Caroff, Bolinsson, and
  Johansson]{caroff2011}
Caroff,~P.; Bolinsson,~J.; Johansson,~J. Crystal {{Phases}} in {{III}}--{{V
  Nanowires}}: {{From Random Toward Engineered Polytypism}}. \emph{IEEE J.
  Select. Topics Quantum Electron.} \textbf{2011}, \emph{17}, 829--846\relax
\mciteBstWouldAddEndPuncttrue
\mciteSetBstMidEndSepPunct{\mcitedefaultmidpunct}
{\mcitedefaultendpunct}{\mcitedefaultseppunct}\relax
\EndOfBibitem
\bibitem[Paladugu \latin{et~al.}(2009)Paladugu, Zou, Guo, Zhang, Joyce, Gao,
  Tan, Jagadish, and Kim]{paladugu2009}
Paladugu,~M.; Zou,~J.; Guo,~Y.~N.; Zhang,~X.; Joyce,~H.~J.; Gao,~Q.;
  Tan,~H.~H.; Jagadish,~C.; Kim,~Y. Evolution of Wurtzite Structured {{GaAs}}
  Shells around {{InAs}} Nanowire Cores. \emph{Nanoscale Res. Lett.}
  \textbf{2009}, \emph{4}, 846--849\relax
\mciteBstWouldAddEndPuncttrue
\mciteSetBstMidEndSepPunct{\mcitedefaultmidpunct}
{\mcitedefaultendpunct}{\mcitedefaultseppunct}\relax
\EndOfBibitem
\bibitem[Lewis \latin{et~al.}(2017)Lewis, Nicolai, K{\"u}pers, Ramsteiner,
  Trampert, and Geelhaar]{Lewis2017}
Lewis,~R.~B.; Nicolai,~L.; K{\"u}pers,~H.; Ramsteiner,~M.; Trampert,~A.;
  Geelhaar,~L. Anomalous Strain Relaxation in {{Core}}\textendash{{Shell}}
  Nanowire Heterostructures via Simultaneous Coherent and Incoherent Growth.
  \emph{Nano Lett.} \textbf{2017}, \emph{17}, 136--142\relax
\mciteBstWouldAddEndPuncttrue
\mciteSetBstMidEndSepPunct{\mcitedefaultmidpunct}
{\mcitedefaultendpunct}{\mcitedefaultseppunct}\relax
\EndOfBibitem
\bibitem[Guo \latin{et~al.}(2013)Guo, Burgess, Gao, Tan, Jagadish, and
  Zou]{Guo2013d}
Guo,~Y. Y.-N.; Burgess,~T.; Gao,~Q.; Tan,~H.~H.; Jagadish,~C.; Zou,~J.
  Polarity-Driven Non-Uniform Composition in {{InGaAs}} Nanowires. \emph{Nano
  Lett.} \textbf{2013}, \emph{13}, 5085--9\relax
\mciteBstWouldAddEndPuncttrue
\mciteSetBstMidEndSepPunct{\mcitedefaultmidpunct}
{\mcitedefaultendpunct}{\mcitedefaultseppunct}\relax
\EndOfBibitem
\bibitem[Heiss \latin{et~al.}(2013)Heiss, Fontana, Gustafsson, W{\"u}st, Magen,
  O'Regan, Luo, Ketterer, {Conesa-Boj}, Kuhlmann, Houel, {Russo-Averchi},
  Morante, Cantoni, Marzari, Arbiol, Zunger, Warburton, and {Fontcuberta i
  Morral}]{Heiss2013a}
Heiss,~M. \latin{et~al.}  Self-Assembled Quantum Dots in a Nanowire System for
  Quantum Photonics. \emph{Nat. Mater.} \textbf{2013}, \emph{12},
  439--444\relax
\mciteBstWouldAddEndPuncttrue
\mciteSetBstMidEndSepPunct{\mcitedefaultmidpunct}
{\mcitedefaultendpunct}{\mcitedefaultseppunct}\relax
\EndOfBibitem
\bibitem[Treu \latin{et~al.}(2013)Treu, Bormann, Schmeiduch, D{\"o}blinger,
  Matich, Wiecha, Saller, Mayer, Bichler, Amann, Finley, Abstreiter,
  Koblm{\"u}ller, and Mork{\"o}tter]{Treu2013}
Treu,~J.; Bormann,~M.; Schmeiduch,~H.; D{\"o}blinger,~M.; Matich,~S.;
  Wiecha,~P.; Saller,~K.; Mayer,~B.; Bichler,~M.; Amann,~M.-C.~C.;
  Finley,~J.~J.; Abstreiter,~G.; Koblm{\"u}ller,~G.; Mork{\"o}tter,~S. Enhanced
  Luminescence Properties of {{InAs}}-{{InAsP}} Core-Shell Nanowires.
  \emph{Nano Lett.} \textbf{2013}, \emph{13}, 6070--7\relax
\mciteBstWouldAddEndPuncttrue
\mciteSetBstMidEndSepPunct{\mcitedefaultmidpunct}
{\mcitedefaultendpunct}{\mcitedefaultseppunct}\relax
\EndOfBibitem
\bibitem[Yuan \latin{et~al.}(2015)Yuan, Caroff, Wang, Guo, Wang, Jackson,
  Smith, Tan, and Jagadish]{Yuan2015b}
Yuan,~X.; Caroff,~P.; Wang,~F.; Guo,~Y.; Wang,~Y.; Jackson,~H.~E.;
  Smith,~L.~M.; Tan,~H.~H.; Jagadish,~C. Antimony Induced \{112\}{{A}} Faceted
  Triangular {{GaAs}}(1-x){{Sb}}(x)/{{InP Core}}/{{Shell}} Nanowires and Their
  Enhanced Optical Quality. \emph{Adv. Funct. Mater.} \textbf{2015}, \emph{25},
  5300--5308\relax
\mciteBstWouldAddEndPuncttrue
\mciteSetBstMidEndSepPunct{\mcitedefaultmidpunct}
{\mcitedefaultendpunct}{\mcitedefaultseppunct}\relax
\EndOfBibitem
\bibitem[Geelhaar \latin{et~al.}(1999)Geelhaar, Marquez, Jacobi, Kley,
  Ruggerone, and Scheffler]{Geelhaar1999}
Geelhaar,~L.; Marquez,~J.; Jacobi,~K.; Kley,~A.; Ruggerone,~P.; Scheffler,~M. A
  Scanning Tunneling Microscopy Study of the {{GaAs}}(112) Surfaces.
  \emph{Microelectron. J} \textbf{1999}, \emph{30}, 393--396\relax
\mciteBstWouldAddEndPuncttrue
\mciteSetBstMidEndSepPunct{\mcitedefaultmidpunct}
{\mcitedefaultendpunct}{\mcitedefaultseppunct}\relax
\EndOfBibitem
\bibitem[Bastiman \latin{et~al.}(2016)Bastiman, K{\"u}pers, Somaschini, and
  Geelhaar]{Bastiman2016}
Bastiman,~F.; K{\"u}pers,~H.; Somaschini,~C.; Geelhaar,~L. Growth Map for
  {{Ga}}-Assisted Growth of {{GaAs}} Nanowires on {{Si}}(111) Substrates by
  Molecular Beam Epitaxy. \emph{Nanotechnology} \textbf{2016}, \emph{27},
  095601\relax
\mciteBstWouldAddEndPuncttrue
\mciteSetBstMidEndSepPunct{\mcitedefaultmidpunct}
{\mcitedefaultendpunct}{\mcitedefaultseppunct}\relax
\EndOfBibitem
\bibitem[K{\"u}pers \latin{et~al.}(2018)K{\"u}pers, Lewis, Tahraoui, Matalla,
  Kr{\"u}ger, Bastiman, Riechert, and Geelhaar]{Kupers2018}
K{\"u}pers,~H.; Lewis,~R.~B.; Tahraoui,~A.; Matalla,~M.; Kr{\"u}ger,~O.;
  Bastiman,~F.; Riechert,~H.; Geelhaar,~L. Diameter Evolution of Selective Area
  Grown {{Ga}}-Assisted {{GaAs}} Nanowires. \emph{Nano Res.} \textbf{2018},
  \emph{1}, 1--9\relax
\mciteBstWouldAddEndPuncttrue
\mciteSetBstMidEndSepPunct{\mcitedefaultmidpunct}
{\mcitedefaultendpunct}{\mcitedefaultseppunct}\relax
\EndOfBibitem
\end{mcitethebibliography}

\end{document}